\documentclass[aps,prb,reprint,superscriptaddress]{revtex4-2}
\usepackage{amssymb}
\usepackage{mathrsfs}
\usepackage{graphicx}
\usepackage[normalem]{ulem}
\usepackage{color}
\usepackage{bm}
\usepackage{physics}
\usepackage{amsfonts}
\usepackage{lipsum}
\newcommand{\GJ}{\textcolor{black}}

\begin{document}


\title{\textbf{Bulk-edge coupling induced by a moving impurity}}

\author{Baikang Yuan}
\affiliation{Centre for Quantum Technologies, National University of Singapore, Singapore 117543, Singapore}
\author{Jiangbin Gong}
\affiliation{Department of Physics, National University of Singapore, Singapore 117551, Singapore}
\affiliation{Centre for Quantum Technologies, National University of Singapore, Singapore 117543, Singapore}
\date{\today}

\begin{abstract}
\GJ{More physics at the boundaries of a topological lattice remains to be explored for future applications of topological edge states.  This work investigates the stability of topological edge states in the presence of a moving impurity. By modeling the impurity as a moving Gaussian on-site potential at the boundary of a two-dimensional (2D) lattice, we show that a moving impurity may cause significant modifications to edge transport, a feature markedly different from the expected robustness of edge transport against a static impurity. We further identify an interesting mechanism to explain the bulk-edge coupling using a co-moving frame, where the density of the bulk states and the degeneracy between the edge states and the bulk become key elements.  The physical insights developed in this work are validated across multiple systems, including Chern insulators, quantum spin Hall insulators, and  Floquet Chern insulators.  Results presented in this work are complementary to our current understanding of the robustness of topological edge transport}.
\end{abstract}

\maketitle

\section{\label{sec:intro}Introduction}
The seminal discovery of the Thouless-Kohmoto-Nightingale-Nijs (TKNN) formula \cite{TKNN}, which connects the Hall conductance with the first Chern number, unveiled the topological origin of the integer quantum Hall conductance \cite{zhang_experiment, pepper_experiment}. Since then, research on topological matter has remained a central focus in condensed matter physics.  In particular, the non-local topological invariants serve as the order parameters in characterizing different quantum phases. Typically defined from the bulk properties of a system, topological invariants are closely related to the presence of robust edge states at the open boundaries of a topological system. The relationship between topological invariants and the topological edge states is known as the bulk-boundary correspondence (BBC) \cite{asboth}. With BBC, one can adiabatically remove an impurity along the system's boundary, resulting in a clean edge that continues to support a topologically protected state. The robustness of these edge states arises from two main aspects: First, band gap protection ensures that edge states, whether chiral or helical, remain localized at the boundary and are difficult to excite into the bulk. These states can be immune to backscattering, enabling nearly perfect transport, \GJ{thus making them} promising candidates for advanced quantum technologies, such as low-loss spintronics \cite{chang_spin, spin_1, spin_2, spin_3}, and advanced sensors \cite{sensing_1,sensing2}. Second, symmetry protection plays a critical role. Perturbations that preserve the system's symmetry cannot easily eliminate the edge states. 

\GJ{To date, the impact of static impurities on a topological lattice is well understood. 
\GJ{On one hand, disorders can renormalize the system parameters and result in topological phases like topological Anderson insulators \cite{TAI_1, TAI_2, TAI_3, TAI_4}, on the other hand, perturbative impurities can impact the edge transport without effecting the bulk topology \cite{dephasing, entanglement_transport, helical_liquid}. }Physical consequences of impurity or noise at the edge of a topological system, as manifested in the quantization of conductance in topological insulators, were also of wide interest \cite{noise1,noise2}.  Much less is known about the implications of an impurity moving on the boundary for edge transport. Related to this question, recent studies have revealed intriguing dynamical effects: mobile impurities can nucleate zero-energy bound states that effectively redefine system boundaries \cite{moving_1}, Majorana domain walls exhibit intrinsic speed limits \cite{moving_2}, and driven magnetic impurities modify helical edge state transmission \cite{moving_3}. Motivated by these previous studies, we address here the following two related questions: (i) How does edge-state dynamics respond to a single moving on-site impurity? (ii) How is the gap protection for topological edge transport threatened by the motion of an impurity?  Further, what difference can the symmetry of a moving impurity make in terms of the induced bulk-edge coupling?}

In order to answer the questions raised above, we extensively study the quantum dynamics of a topological lattice with a moving impurity and observe that a moving impurity indeed induces considerable loss of edge state population.  {This indicates that {the energy scale associated with a moving impurity can considerably harm the gap protection normally assumed for topological edge states}. To explain the underlying physics, we introduce a frame co-moving with the impurity, and there the energy of the unperturbed edge states can become, depending on the velocity of the impurity, highly degenerate with the bulk.  Because the impurity viewed in the co-moving frame, though static again, can be viewed as a spatially complicated perturbation, the density of the bulk states on {resonance} with the edge states becomes a key quantity to understand the population loss from edge state dynamics.}  Such physical insights are validated across multiple systems, including Chern insulators, 2D quantum spin Hall insulators, and Floquet Chern insulators. 

The manuscript is structured as follows. In Sec.~\ref{sec:model}, we describe our simulation setup and present dynamics simulation results, followed by our analysis by a simple perturbation theory analysis.  In Sec.~\ref{sec:dis}, we extend our investigations to a non-equilibrium topological phase model. Results therein further support our understanding.  Finally, we summarize and discuss our findings in Sec.~\ref{sec:con}.

\section{\label{sec:model}setup and results}
\subsection{Chiral edge states}
We begin by investigating chiral edge states in a Chern insulator — a paradigmatic two-dimensional system lacking anti-unitary symmetries. \GJ{ There} topological protection arises solely from the bulk band gap. In this case without any symmetry, the BBC ensures robust edge transport as long as the band gap remains open.  
Applying this expectation to our setting, a localized impurity should have little impact on the quantum transport along the edge. Our task here is to look into how the situation changes qualitatively and quantitatively when the impurity is set into motion. 

Consider then the Qi-Wu-Zhang  (QWZ) model \cite{asboth} on a square lattice, with the Bloch Hamiltonian
\begin{equation}
\begin{aligned}
H_{\text{QWZ}}(k_{x},k_{y},m) &= J\sin(k_{x})\sigma_{x} + \sin(k_{y})\sigma_{y} \\
&\quad + \left[m + \cos(k_{x}) + \cos(k_{y})\right] \sigma_{z}.
\end{aligned}
\end{equation}
Unless specified otherwise to tune the edge state dispersion relation, we always assume $J=1$. In the regime of $|m|<2$, this Chern insulator supports chiral edge states. $k_x\in[0, 2\pi]$ and $k_y\in[0, 2\pi]$ here are quasi-momenta along the $x$ and $y$ dimentions. 

In our considerations below, we apply the open boundary condition in the $y$ dimension and the periodic boundary condition in the $x$ dimension.  We examine the time evolution of an initial state at the $y=0$ edge, denoted $\ket{\Phi(0)}$, by considering a Gaussian wave packet moving along the $x$ direction
\begin{equation}
    \ket{\Phi(0)} = \frac{1}{\cal N} \sum_{k_x} e^{- (k_x - k_0)^2/2\alpha^2} e^{-iN_c(k_x - k_0)}e^{ik_x n} \ket{\phi_{k_x}},
\end{equation}
where $\ket{\phi_{k_x}}$ is the gapless edge state at momentum $k_x$ and $n$ is unit cell index along $x$, $k_0$ is the central momentum of this wave packet as a superposition of many quasi-momentum states,  ${\cal N}$ is the normalization factor, $\alpha$ controls the wavepacket width, and $N_c$ is the central position of the wave packet.  Such a Gaussian wave packet in momentum space remains a Gaussian wave packet in real space under the Fourier transformation. In our numerical simulations presented below, $k_0$ will be chosen as the central momentum possessed by the in-gap edge states.  

\begin{figure}[htbp]
    \centering
    \includegraphics[scale=0.6]{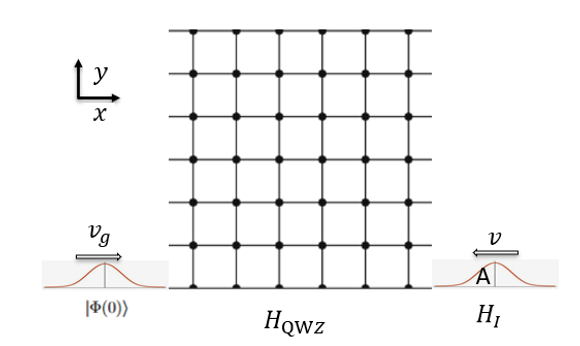}
    \caption{
    Schematic of our dynamics simulation setup. The topological lattice is described by the QWZ model with a periodic boundary along $x$ (length $N_x=41$) and an open boundary along $y$ (width $N_y=21$). The initial state is a Gaussian wave packet localized on the bottom edge ($y=0$), with a group velocity $v_g$.  A Gaussian-shaped on-site impurity is moving with velocity $v$ from right to left.}  
    \label{setup}
\end{figure}

We next introduce a mobile impurity as a moving Gaussian on-site potential
\begin{equation}
    H_{I}(x, y=0,t) = A  e^{-\frac{1}{2} [x - (x_0 - vt)]^2} \sigma_{z} ,
\end{equation}
where $A$ is the amplitude of the onsite potential. $x_0$ is the central position of the potential and $v$ is the velocity of the impurity. The choice of the operator $\sigma_z$ acting on the internal degrees of freedom here is not essential because the Chern insulator does not require any symmetry protection. As an example, here we use opposite on-site potentials on the two sublattices. Fig.~\ref{setup} shows a schematic of our setup.

We are now ready to investigate the robustness of edge states in terms of the dynamics of the launched initial state $\ket{\Phi(0)}$  under the full Hamiltonian
\begin{equation}
    H(t) = H_{\text{QWZ}} + H_I(t).
\end{equation}
The population of the time evolving state still remaining at the $y=0$ bottom edge is given by
\begin{equation}
    A_e(t) = \sum_{(x,y);\ y\in \text{edge}} |\phi(x,y,t)|^2. 
\end{equation}
The summation in the equation above is over all sites with different index $x$ but with $y=0$. The population loss of the edge transport during the time evolution is hence given by
\begin{equation}
    L(t) = A_e(0) - A_e(t). 
\end{equation}
This loss is a quantitative measurement of the bulk-edge coupling, as the population loss is due to scattering into the bulk.

\subsection{The co-moving frame}
To digest our dynamics results, we further prepare our readers by introducing a frame co-moving with the impurity. As seen below, this will be highly useful in developing qualitative insights into the dynamics. To that end, we approximately view the above-introduced QWZ lattice Hamiltonian as a continuous system (that is, treating the lattice index as a continuous variable and also viewing the quasi-momentum variables $k_x$ and $k_y$ as normal momentum variables). In this continuum picture, the total Hamiltonian, including the moving impurity, becomes
\begin{equation}
\begin{aligned}
    H_{\text{continuum}} &= \int_{k_x,k_y} H_{\text{QWZ}}(k_x,k_y)\ket{k_x,k_y}\bra{k_x,k_y} +\\
    &\int_{x,y=0} A e^{-\frac{1}{2}[x - (x_0 - vt)]^2}\sigma_z\ket{x,y}\bra{x,y}.
\end{aligned}
\end{equation}
This treatment then allows us to perform a Galilean transformation to the impurity's rest frame via the unitary operator
\begin{equation}
    U(t) = e^{ivt\hat{k_x}\otimes\sigma_0},
\end{equation}
where $\hat{k_x} = -i \partial_x$ is now viewed as the position shift operator. The transformed Hamiltonian becomes
\begin{equation}
\begin{aligned}
    H' &= U^{\dagger}(t) H_{\text{continuum}} U(t) - iU^{\dagger}\frac{d}{dt}U\\
    &= \int_{k_x,k_y} H_{\text{QWZ}}(k_x,k_y)\ket{k_x,k_y}\bra{k_x.k_y} \\
    &+ \int_{x,y=0} A e^{-\frac{1}{2}(x - x_0)^2}\sigma_z\ket{x,y}\bra{x,y} \\
    &+ \int_{k_x} k_xv\ket{k_x}\bra{k_x}.
\end{aligned}
\end{equation}
Finally, we use the above Hamiltonian to arrive at a lattice model in the co-moving frame, with mixed boundary conditions as in Fig.~\ref{setup}. The lattice model Hamiltonian in the co-moving frame is hence obtained as follows: 
\begin{equation}\label{discrete}
\begin{aligned}
    H' &=\sum_{k_x,y}H_{\text{QWZ}}(k_x,y)\ket{k_x,y}\bra{k_x,y} \\
    &+ \sum_{k_x}{k_x v} \ket{k_x}\bra{k_x}\\
    &+ \sum_{x,y=0} A e^{-\frac{1}{2}(x - x_0)^2}\sigma_z\ket{x,y}\bra{x,y}.\\ 
\end{aligned}
\end{equation}
The first term of $H'$ is nothing but the same QWZ Hamiltonian under the mixed boundary condition and hence is a function of $k_x$ and the lattice index along the $y$ dimension. $H'$ also acquires a $k_x$-dependent energy shift $k_xv$. The last term in the above equation now represents a spatial impurity, now being static as constructed.  This motivates us to combine the first two terms of $H'$ together as a topological system $H_{\rm co}^{0} = H_{\text{QWZ}} + k_xv$. The remaining static impurity potential, namely,  $V_{\rm co}= \sum_{x, y=0} A e^{-\frac{1}{2}(x - x_0)^2}\sigma_z\ket{x,y}\bra{x,y}$ along the $y=0$ edge, can be treated as a perturbation. Using this co-moving frame, the potential impact of the impurity velocity is now all captured into $H^0_{\rm co}$.  Certainly, $H^0_{\rm co}$ is rather simple only in the momentum space picture adopted here.  In real space, $H^0_{\rm co}$ when transformed to the $x$ representation will assume rather complicated lattice couplings.  


One must not forget that when deriving Eq.~(\ref{discrete}) using the continuum model, we assumed that the response of the system to the moving impurity is local and smooth. A large-amplitude and fast-moving impurity can effectively create sharp and non-smooth responses, and in these cases, our treatment above is not expected to be quantitatively correct. We hence restrict ourselves to cases where the impurity's amplitude and speed are at most moderate, and our perturbation theory below, based on $H_{\rm co}^{0}$ plus a static perturbation, is mainly to digest dynamics simulation results and develop qualitative insights. 

Using the co-moving frame introduced above, let us now treat $V_{\rm co}$ as a static perturbation to $H^0_{\rm co}$. The first-order perturbation theory (PT) yields the following transition probability:
\begin{eqnarray}
\label{perturb}
    P_{mn}(t)&= &\left|\int_0^tdt'e^{i(E_m^0-E_n^0)t'}\langle m |V_{\rm co}|n\rangle\right|^2 \nonumber \\
    &=& |\langle m |V_{\rm co}|n\rangle|^2  \left|\frac{e^{i\omega_{mn}t}-1}{i \omega_{mn}}\right|^2
\end{eqnarray}
where $P_{mn}(t)$ denotes the transition probability from $H^0_{\rm co}$'s eigenstate $n$ to its eigenstate $m$ over a duration from time 0 to time $t$. $\omega_{mn}=E_m^0-E_n^0$ are the transition frequency ($\hbar=1$ is our dimensionless units). Apparently, because the perturbation is static in the co-moving frame, the transitions will favor degeneracy between the initial and final states. One may further expect something similar to the Fermi's golden rule if we assume that the matrix elements $|\langle m |V_{\rm co}|n\rangle|^2$ do not have peculiar features (to be elaborated later). Note that the transition probabilities obtained in the co-moving frame can be directly used to calculate the population loss from the edge, because the unitary transformation $U(t)$ between the two frames is only changing phases along the $x$ dimension -- it does not alter the loss function for an initial state launched along the $y=0$ boundary.

\subsection{Edge state dynamics in the presence of a mobile impurity}
 We examine edge state transport based on a second-order Trotter decomposition \cite{trotter_1, trotter_2} of the lattice Hamiltonian in the lab frame. The convergence of our time evolution results is verified and can be regarded as numerically exact results. Fig.~\ref{l_t} shows the time dependence of the loss function $L(t)$ obtained from our dynamics simulations. The group velocity of the Gaussian wave packet—comprising a superposition of edge states at the $y=0$ edge—is estimated from the edge state at the central momentum {$k_0$ (e.g., for $m>0$, we choose $k_0=\pi$ because this is also the central momentum of all in-gap edge states). The dispersion relation of the edge states for the chosen system parameters can be obtained from the edge Hamiltonian (for details, see Appendix \ref{append_A}): {$H_{\rm edge} \approx J (k_x - \pi) \sigma_x$}. The group velocity of the initial Gaussian wave packet is hence given by $\partial E/ \partial k_x$, yielding $v_g=J=1$. Later, when we need to tune $v_g$, we just choose different coefficients of the $J\sin(k_x)$ term in the QWZ model. 
 
 Fig.~\ref{l_t}(a) depicts the robustness of edge state dynamics against a static impurity. The peak in the loss function as a function of time occurs when there is a maximal overlap between the time-evolving wave packet and the moving impurity -- the encounter between the impurity and the edge state redirects some population away from the $y=0$ edge. However, when the wavepacket and the impurity are spatially separated again, the loss function is back to almost zero, indicating the penetration of some population into the bulk is released back to the edge, with a residual population loss of about $1\%$ attributed to finite-size effects in our simulations. The second peak of the loss function in Fig.~\ref{l_t}(a) is only because of the periodic boundary condition along the $x$-dimension, so that the moving impurity is allowed to collide with the wavepacket again after having explored the entire $y=0$ edge.   
 
Interestingly, the behavior of the loss function is drastically different when the impurity is mobile, as shown in Fig.~\ref{l_t}(b). Here, the population loss increases sharply during each collision between the time-evolving wave packet and the moving impurity. There is essentially no population backflow from the bulk to the edge. The fair agreement between the exact time evolution (TE) simulations and the perturbation theory (PT) indicates that even for $A=0.5$, the perturbative treatment is still useful (as confirmed by more results below). The marked difference in bulk-edge coupling caused by a moving impurity motivates us to develop more understanding of the underlying physics. Below, we propose to use the population loss during the first encounter with the impurity as a key indicator to guide our further analysis.  

\begin{figure}[htbp]
    \centering
    \includegraphics[scale=0.255]{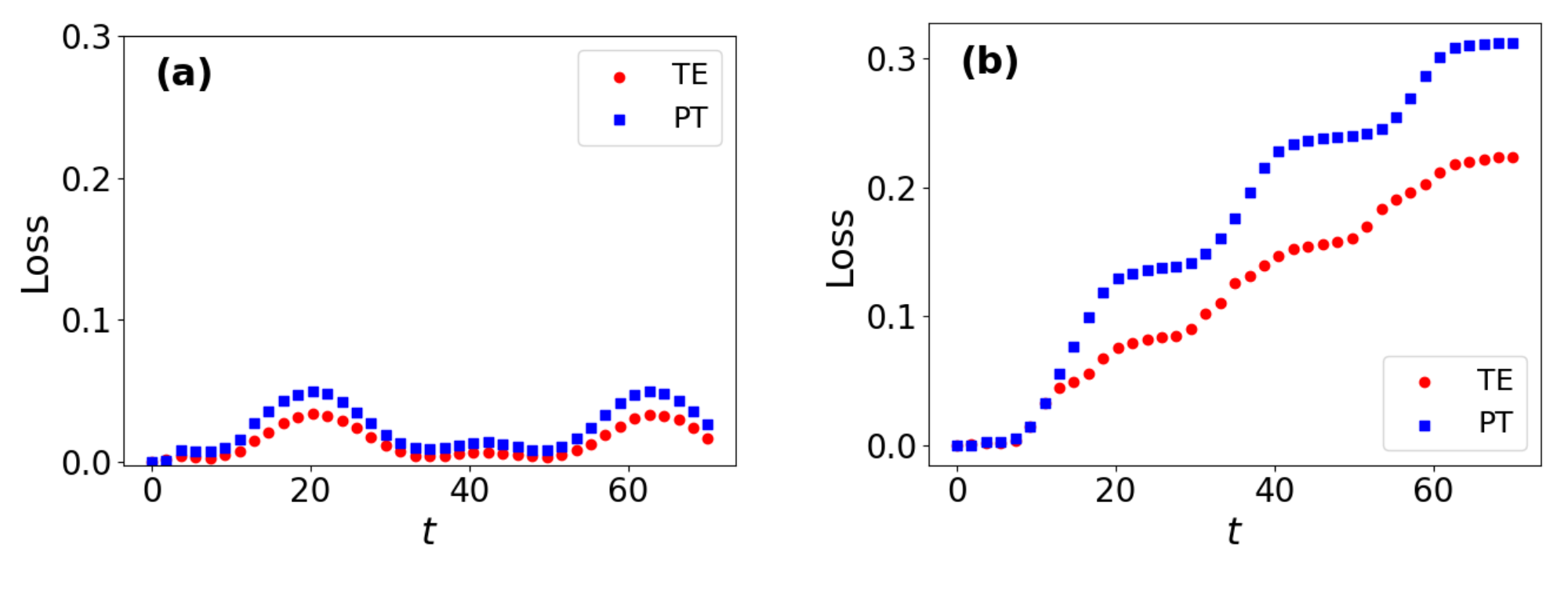}
    \caption{Population loss to the bulk as a function of time, with a static (a) and moving (b) impurity. Other system parameters are $A=0.5, v=1$. Here and in the following figures, when presenting the loss function, circles represent time evolution (TE) simulations, whereas squares represent results from perturbation theory (PT). Other system parameters are $m=1.5, J=1$.}
    \label{l_t}
\end{figure}


\subsection{Population loss due to edge-bulk coupling}
We now examine how the loss function depends on features of the impurity and on other system parameters. For a moving impurity, the key parameters are its strength $A$ and velocity $v$. On the otherwise clean topological system side, the key parameters are the edge-state group velocity $v_g$ and the bulk band gap $\Delta$.  

\begin{figure}[htbp]
    \centering
    \includegraphics[scale = 0.255]{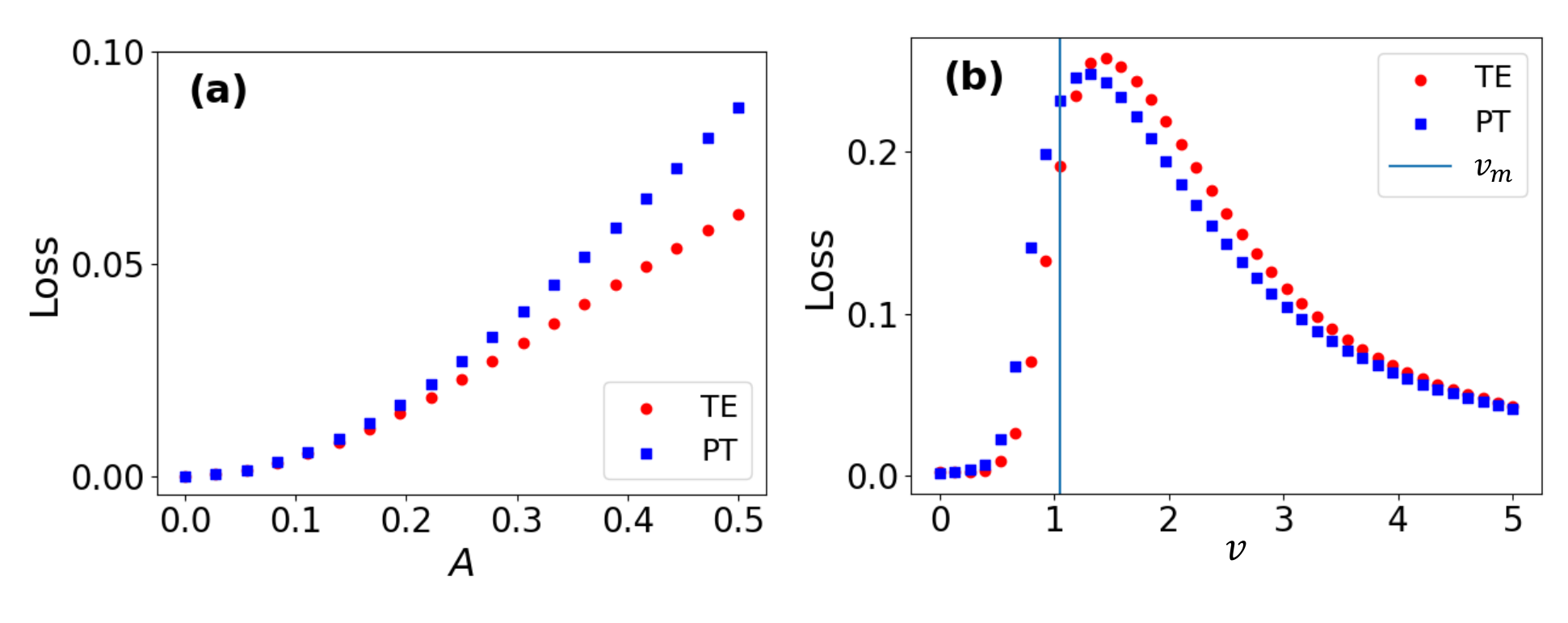}
    \caption{(a) Probability loss from edge to bulk states versus the impurity strength $A$, with fixed $v=1$. (b) The same loss function versus the velocity $v$ of the moving impurity, with fixed impurity strength $A=0.5$. The vertical line indicates a special velocity where there is the maximal density of the bulk states at a reference energy of the initial state.  Other system parameters are $m = 1, J = 1$.}
    \label{v_A}
\end{figure}

As shown in Fig.~\ref{v_A}(a), from our dynamics simulations we observe an approximately monotonic relationship between the loss and the impurity strength $A$. This is expected because stronger impurities should create larger bulk-edge coupling. The agreement between the actual dynamics simulations and the first-order perturbation theory for small $A$, as shown in Fig.~\ref{v_A}(a), confirms our perturbative treatment elaborated above. In terms of how the loss function depends on the velocity of the moving impurity,   Fig.~\ref{v_A}(b) compares results from dynamics calculations and from the perturbation theory for a relatively large impurity strength $A=0.5$.  Fair agreement is also observed, albeit with some expected discrepancy for such a strong impurity strength. Later, we shall discuss the vertical line also plotted in Fig.~\ref{v_A}(b). 

To investigate how the bulk-edge coupling depends on the dispersion relation on the edge, we now allow to tune the system parameter $J$, which proportionally determines the group velocity $v_g$ of the edge states. In particular, the edge Hamiltonian of our QMZ model reads $H_{\text{edge}} = J (k_x-\pi) \sigma_x$, with the dispersion relation {$E_{\rm edge} = \pm J(k_x-\pi)$}. The group velocity of the edge state is hence given by $v_g = \pm J$. For later reference, we also mention that the effective Hamiltonian that describes the edge state in the co-moving frame is then given by $H_{\text{eff}} = J (k_x-\pi)\sigma_x + v k_x$, where the second term comes from the co-moving frame transformation. In addition,  we also tune the system parameter $m$ to control the bulk gap denoted $\Delta$. 

\begin{figure}[htbp]
    \centering
    \includegraphics[scale=0.255]{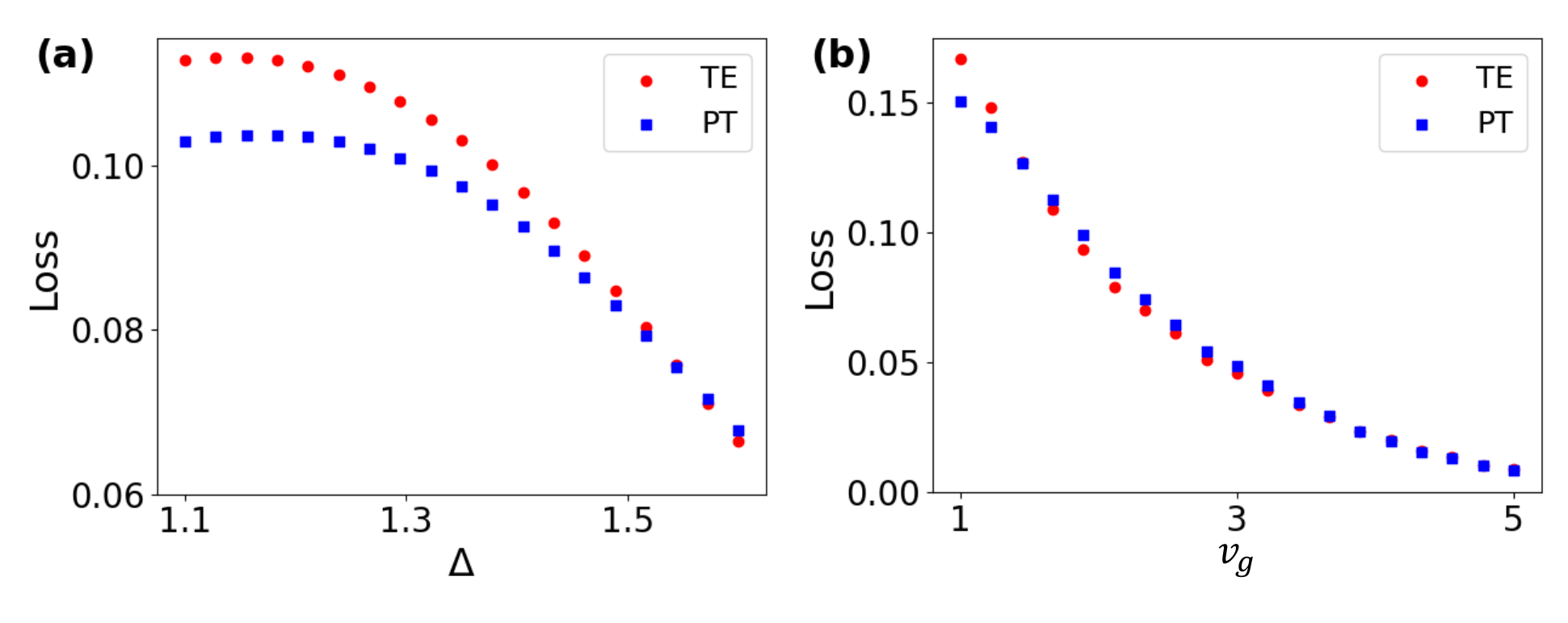}
    \caption{(a) Probability loss function versus the band gap with fixed $v_g=1$, with the explicit expression for the bulk gap given by $\Delta = 2 |m|$ for $m\in [0.5, 1]$. (b) Loss function versus the edge state group velocity $v_g$ with fixed band gap $\Delta=2$. Unless otherwise specified, other system parameters are fixed at $m=1$, $A=0.5$, $J=1$ in all subsequent plots.}
    \label{loss_system}
\end{figure}

The loss function exhibits a monotonic dependence on the band gap $\Delta$, as shown in Fig.~\ref{loss_system}(a). A large band gap enhances the energy separation between bulk and edge states, suppressing excitations from edge to bulk – a behavior fully consistent with the concept of band gap protection. Furthermore, Fig.~\ref{loss_system}(b) reveals that the loss function is inversely proportional to the group velocity of the edge states in the large $v_g$ regime: faster-moving edge states (larger group velocity) experience reduced interaction times with the impurity, leading to lower excitation probabilities. This observation is well aligned with established theoretical predictions \cite{group_1, group_2} regarding the group velocity dependence of topological edge state resistance in real materials, where higher propagation velocities typically correspond to improved transport characteristics.  Notably again, it is seen from  Fig.~\ref{loss_system}(a) and  Fig.~\ref{loss_system}(b) that our simple perturbation treatment stays very useful. 

Summarizing all our observations above, the most interesting aspect of the loss function is captured in Fig.~\ref{v_A}(b): there is a pronounced peak of the loss function as the velocity of the moving impurity increases. The good performance of our perturbative treatment so far also motivates us to physically explain why there should be a special velocity yielding the maximal population loss arising from bulk-edge coupling.

\subsection{The impurity velocity that yields the  maximal population loss}
To understand the peak in Fig.~\ref{v_A}(b), we return to the perturbation theory result in Eq.~(\ref{perturb}).  Because the perturbation treated in the co-moving frame is static, the population will leak to the bulk states that share the same energy as the initial state launched along the edge, so that $\omega_{mn} = E_m - E_n$ is almost zero (spectral degeneracy between the edge and the bulk). Note also that the matrix elements of the static perturbation $\langle m|V_{\rm co}|n\rangle$ are rather complicated. One may argue that these matrix elements can be roughly treated as a random variable. This being the case, the total population loss from the edge, within the perturbation theory perspective and analogous to the Fermi golden rule, will be proportional to the density of states of the bulk evaluated at the energy of the initial state. Taking into account both spectral degeneracy and the density of bulk states, we hence infer that the maximal population loss occur when there is the maximal spectral degeneracy between the edge state and the bulk states. {Below and also in Fig.~3(b), the velocity of the moving impurity that yields such maximal spectral degeneracy will be denoted $v_m$.}

Regarding the density of states (DOS) of the bulk states in the co-moving frame, evaluated at the energy of the initial state launched at the edge, we operationally extract its values using the  Gaussian broadening:
$ \text{DOS} = \sum_i \delta(E_i - E) \approx \sum_i e^{- \frac{(E_i - E)^2}{\eta^2}}$,
where $\eta$ controls the energy resolution (we will choose $\eta=0.5$ throughout). The reference energy $E$ is taken as that of the central momentum $k_0$ of the Gaussian wave packet launched initially,  which is $\pi v$ in the co-moving frame. The maximal spectral degeneracy arises at the velocity $v_m$ that maximizes the DOS at this reference energy.

\begin{figure}[htbp]
    \centering
    \includegraphics[scale=0.255]{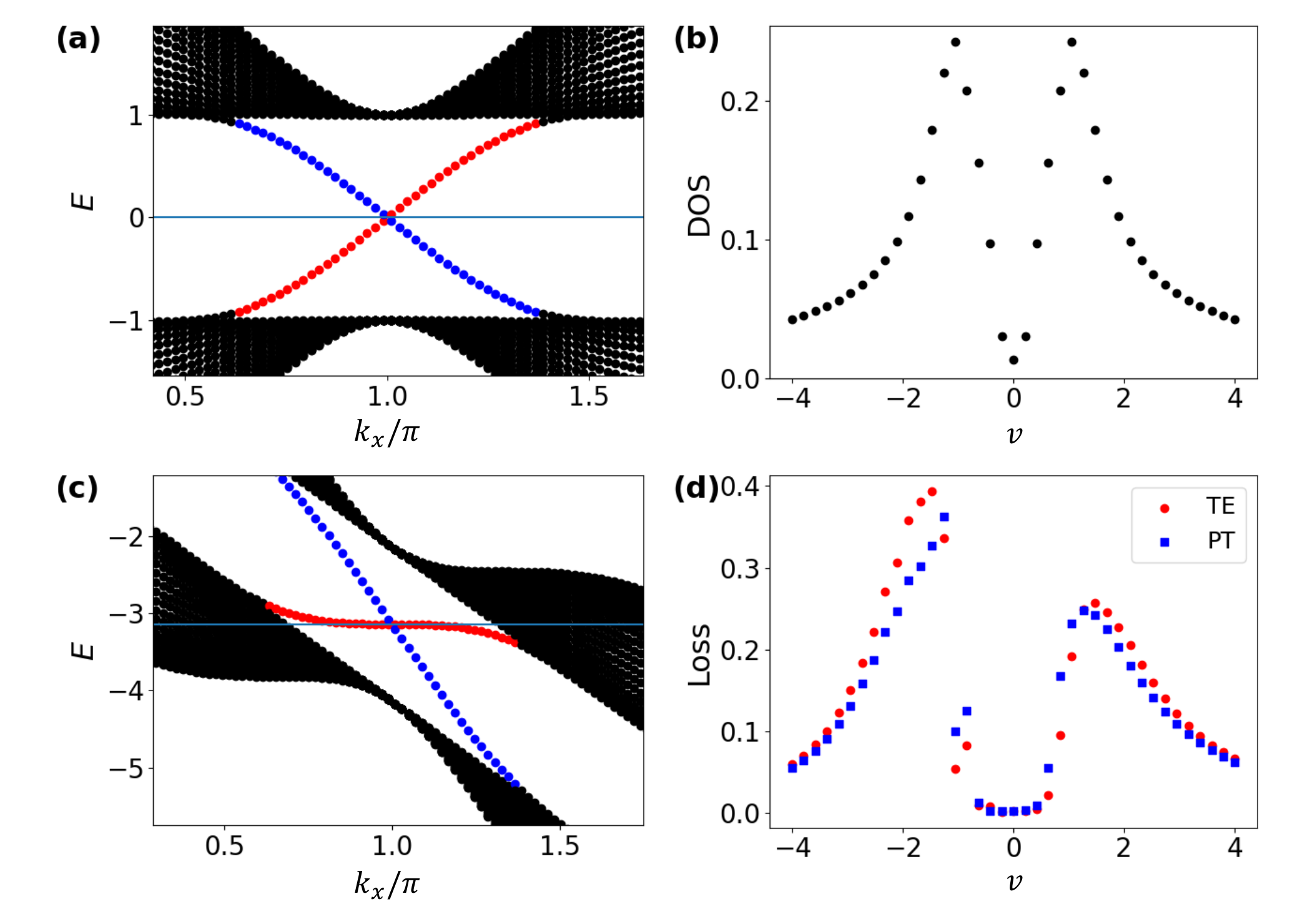}
    \caption{Topological band spectrum with $v=0$ in (a), and in the co-moving frame with $v=-1$ in (c). The black bands indicate the bulk states, and red and blue lines indicate a pair of chiral edge states. The horizontal line is the reference energy used in calculating the DOS. (b) The DOS of the bulk states in the co-moving frame is evaluated at the energy of the central momentum of the initial Gaussian state. (d) Loss function versus the impurity velocity $v$.}
    \label{spectrum}
\end{figure}

With the insights developed above,  we now examine in Fig.~\ref{spectrum} again the population loss from the edge vs the velocity of the moving impurity, in connection with the bulk band plots in the original frame and in the co-moving frame, as well as the dependence of the density of bulk states evaluated at the energy of the initial state. Special attention should be paid to the bulk bands in the co-moving frame, where our perturbation theory can be best digested.  At $v=-1$, the edge state dispersion (always launched along the red branch) in the co-moving frame [panel (c)] essentially becomes horizontal, thus almost all the momentum components of the initial Gaussian wavepacket can ``cut" through all available bulk states along the horizontal direction. In this configuration, the initial state is expected to naturally possess maximal spectral degeneracy with the bulk states.   This is confirmed by the plot of the density of bulk states in Fig.~\ref{spectrum}(b), where it reaches its maximum also at around $v=-1$.  Indeed, in Fig.~\ref{spectrum}(d), it is seen that the maximal loss of population indeed occurs at around $v=-1$.  Interestingly, the population loss depicted in  Fig.~\ref{spectrum}(d) is not symmetric with respect to $v$.  This can be qualitatively explained as follows.  For the regime of positive $v$ (which is also presented in Fig.~3), at around $v=1$ the edge state of the blue branch has almost horizontal dispersion in the co-moving frame. However, because the initial state is on the red branch, only the component around the central momentum $k_0$ has a chance to be degenerate with the bulk, and hence suffers from less population loss. This way, we are able to explain how bulk-edge coupling leads to different consequences if the moving impurity and the edge state group velocity are along the same or the opposite direction (corresponding to negative and positive $v$, respectively)




 The flattening of the edge state dispersion in the co-moving frame is a very useful qualitative insight to anticipate the possibility of maximal spectral degeneracy between the initial state and the bulk. However, it is an oversimplified picture when applying this geometric criterion without checking the overall profile of the bulk band.  Let us now investigate a representative case with  $J<1$,  where the energy gap minimum shifts from the high-symmetry point $k_x = \pi$ to two symmetry points at $k_x = \pm \pi/2$, as shown in Fig.~\ref{small_J}(a). This band restructuring also alters the bulk band profile in the co-moving spectrum. As illustrated in Fig.~\ref{small_J}(b), when the edge state dispersion in the co-moving frame is nearly horizontal, it can only ``cut through" a small fraction of the bulk states along the horizontal direction. By contrast, when we further increase the velocity of the impurity in Fig.~\ref{small_J}(c), the initial state (launched at the center of the red branch) can be degenerate with more bulk states. Therefore, in this case, the DOE peaks at a $v$ much different from the one that tends to flatten the edge state dispersion.  Indeed, Fig.~\ref{small_J}(d) shows the actual loss function versus $v$, and it is seen that the loss indeed almost peaks at the maximal DOE of the bulk state evaluated at the initial state energy, not at the velocity that flattens the edge state dispersion in the co-moving frame.  This example hence clearly indicates that the overall shape of the bulk band is also an important aspect in analyzing bulk-edge coupling induced by a moving impurity.   

\begin{figure}[htbp]
    \centering
    \includegraphics[scale=0.245]{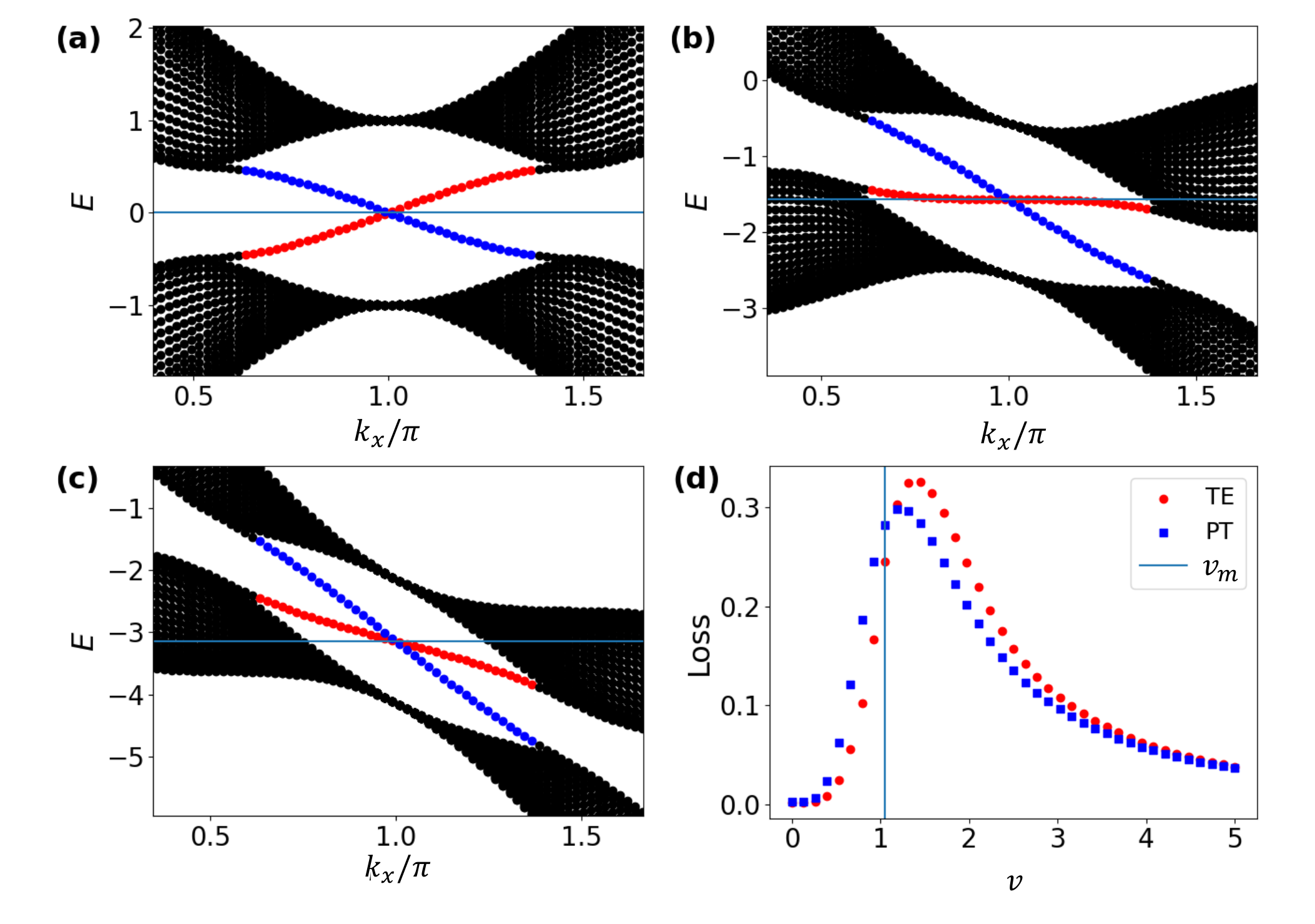}
    \caption{Topological band spectrum of the original system with $v=0$ in (a), and in the co-moving frame with $v=-0.5$ in (b), and in the co-moving frame with $v=-1$ in (c). (d) Loss function versus the velocity of the moving impurity. The vertical line in (d) indicates the velocity determined by having maximal DOS of the bulk in the co-moving frame evaluated at the energy of the central momentum of the initial Gaussian state.  In all the panels here $J=0.5$.}
    \label{small_J}
\end{figure}

\subsection{Helical edge states}
Helical edge states exhibit remarkable robustness against non-magnetic scattering due to spin-momentum locking, thus forbidding back-scattering in the absence of spin-orbit coupling. 
How does a moving impurity in this case affect the robustness of helical edge state transport?
To answer this question, we now look into the Bernevig-Hughes-Zhang (BHZ) model \cite{BHZ} as a specific example, of which the Hamiltonian reads
\begin{equation}
\begin{aligned}
    H = s_0 \otimes [(u + \cos(k_x) + \cos(k_y)) \sigma_z + \sin(k_y) \sigma_y]\\
     + s_z \otimes \sin(k_x) \sigma_x + s_x \otimes \hat{C}.
\end{aligned}
\end{equation}
The operators $s_0$, $s_x$, and $s_z$ here represent Pauli matrices acting on the spin degree of freedom, while $\hat{C}$ describes the coupling between two spin blocks. In our analysis, we set $\hat{C}=0$, yielding a minimal model that captures the quantum spin Hall effect in HgTe quantum wells \cite{BHZ}. This system then reduces to a direct sum of two QWZ models with opposite Chern numbers, making the existence of counter-propagating helical edge states at the boundary highly intuitive. {We have also checked a case with significant coupling between the two spin blocks, namely,  $\hat{C} = 0.3 \sigma_y$, and the results obtained (not shown) are quite similar to the case of $\hat{C}=0$.

We first introduce a non-magnetic impurity potential of the form: $ H_I' = s_0 \otimes H_I$, where $H_I$ is defined in Eq.~(3) and $s_0$ is the identity matrix in the spin space.  Evidently, the impurity term $H_I'$ depends on time $t$, thus breaking the time-reversal symmetry. However, the absence of inter-spin-block coupling preserves protection against back-scattering - the only consequence of bulk-edge coupling due to the moving impurity is hence bulk excitation. In the second scenario, we consider an impurity with spin-orbit coupling (SOC), namely, $H_I'^{\text{SOC}} = s_x \otimes H_I$, which allows transitions between edge channels of different helicity. 

Results of our dynamics simulations and perturbative treatment are shown in Fig.~\ref{helical}, with two key findings. First, when the impurity velocity is close to the group velocity of the edge state, a situation with the maximal spectral degeneracy between the initial state and the bulk in the co-moving frame, we observe the maximal loss of population to the bulk, as evidenced by the peak of the loss function vs the vertical line determined by the maximal spectral degeneracy in Fig.~\ref{helical}(a) and (b). Interestingly, the impurity with SOC causes much less population loss, because it enables backscattering that still confines the wavepacket along the boundary. 

\begin{figure}[htbp]
    \centering
    \includegraphics[scale=0.255]{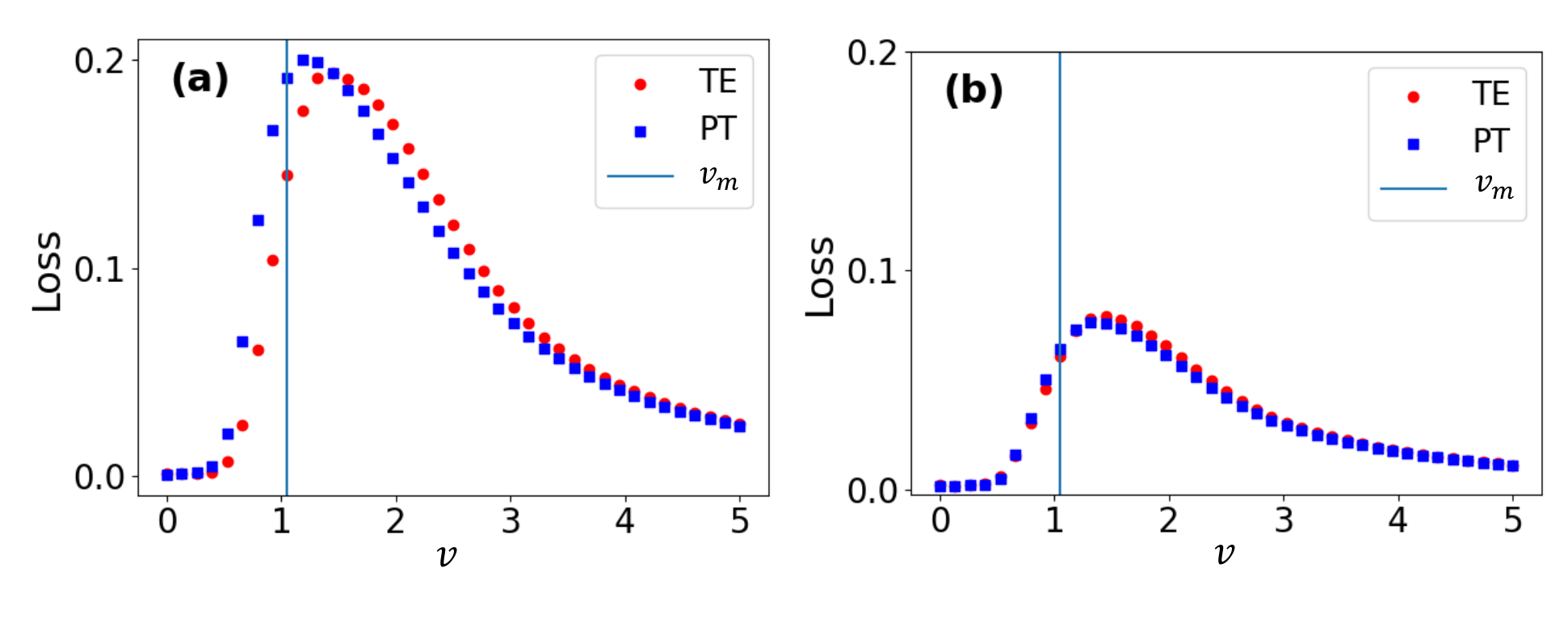}
    \caption{Loss function for helical edge states with a non-magnetic impurity $H_I' = s_0 \otimes H$ in (a) , and with spin-orbit coupling impurity $H_I'^{\text{SOC}} = s_x \otimes H_I$ in (b). The vertical lines in both panels indicate the impurity velocity that maximizes the spectral degeneracy between the bulk and the edge in the co-moving frame.}
    \label{helical}
\end{figure}

\section{\label{sec:dis} Extension to Floquet Chern insulators}

As demonstrated above, in the co-moving frame, it is seen clearly that a moving impurity can damage bulk gap protection by inducing degeneracy between bulk states and edge states. This leads us to investigate whether our general insights apply to the so-called Floquet Chern insulators as non-equilibrium counterparts of static Chern insulator phases.   The underlying subtlety is that in such non-equilibrium systems, energy is no longer conserved.  Nevertheless, below we show that we may still use the effective Floquet Hamiltonian associated with one driving period (which can be used to compute the quasi-energy spectrum) to predict when the maximal spectral degeneracy between the initial state and the bulk states occurs in order to digest the dynamics simulation results.

Let us now consider a Floquet Chern insulator model by periodically quenching the QWZ model, such that the time-dependent Bloch Hamiltonian becomes
\begin{equation}
H(\mathbf{k}, t) = 
\begin{cases}
H_1(\mathbf{k}) = 3\gamma_1 \sigma_x, & T \leq t < T + \frac{T}{3}; \\
H_2(\mathbf{k}) = 3\gamma_2 \sigma_y, & T + \frac{T}{3} \leq t < T + \frac{2T}{3}; \\
H_3(\mathbf{k}) = 3\gamma_3 \sigma_z, & T + \frac{2T}{3} \leq t < 2T,
\end{cases}
\end{equation}
where we set $T=1$ as the driving period, $\gamma_1 = J_1 \sin(k_x)$, $\gamma_1 = J_2 \sin(k_y)$, $\gamma_3 = J_3 [M + \cos(k_x) + \cos(k_y)]$. This model has been carefully studied in previous studies \cite{Umer, recipe}, with the attractive feature that this model can generate an unbounded number of chiral edge state channels by tuning the system parameters. The effective Floquet Hamiltonian for one driving period is given by: $H_{\text{eff}} = -i \log(U)$, where $U = e^{-i\frac{1}{3} H_3} e^{-i\frac{1}{3}H_2} e^{-i \frac{1}{3} H_1}$ is the Floquet operator. Here, we choose the system parameters to ensure that one pair of chiral edge states exists in the zero quasi-energy gap. Fig.~\ref{floquet}(a) depicts the Floquet quasi-energy bulk bands with edge states in the zero quasi-energy gap. We then execute dynamics simulations using a moving impurity under the same boundary conditions as in the above Chern insulator case. Fig.~\ref{floquet}(b) presents the population loss to the bulk as a function of the impurity velocity again. It is seen from Fig.~\ref{floquet}(b) that the loss peaks at a velocity very close to when the maximal spectral degeneracy (based on the effective Floquet Hamiltonian) occurs, thus again showing the usefulness of our physical picture based on a simple perturbative treatment in understanding the actual dynamics simulations.     

\begin{figure}[htbp]
    \centering
    \includegraphics[scale=0.255]{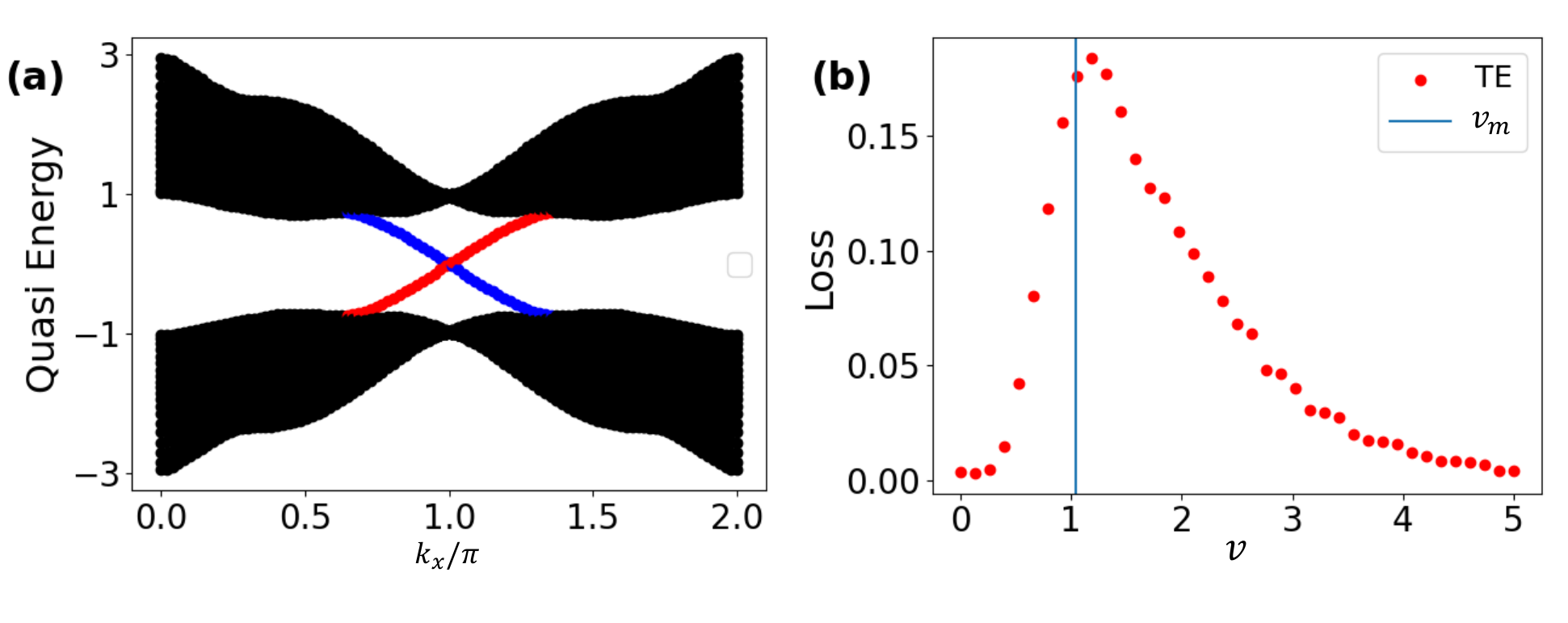}
    \caption{Bulk-edge coupling in the case of a Floquet topological Chern insulator. (a) Quasi-energy spectrum shows chiral edge states (red/blue) in the zero-quasienergy gap, but with no edge states in the $\pi$-quasienergy gap. (b) Loss function (probability to the bulk) versus the velocity of a moving impurity. Other system parameters are chosen as $J_1 = J_2 = J_3 = m = 1$. The vertical line represents the impurity velocity that yields the maximal spectral degeneracy with the bulk states evaluated at the energy (using the Floquet effective Hamiltonian) of the initial Gaussian wavepacket launched at the center of the red branch of the edge states in panel (a). }
    \label{floquet}
\end{figure}

\section{\label{sec:con}conclusion}
This work is concerned with how topological gap protection may be threatened by a moving impurity along the edge of a two-dimensional topological lattice.  The main finding is that the bulk-edge coupling induced by a moving impurity will lead to a sharp velocity-dependent population loss from the system's boundary to the bulk.  By working with a frame co-moving with the impurity, we are able to use a simple perturbation theory to digest the impurity-velocity dependence of bulk-edge coupling. We show that the density of bulk states in the co-moving frame that are degenerate with the initial state launched at the edge is the main determining factor to digest and predict the features of the population loss. This also enables us to explain why a head-on collision between an edge state and a moving impurity can induce less population loss than in a rear-end collision.  Interestingly, the overall bulk spectrum profile and the whereabouts of the in-gap edge states are both important aspects in bulk-edge coupling because they will both affect the density of the bulk states in the co-moving frame. We have also successfully applied our physical insights to non-equilibrium Chern insulator phases where the concept of energy degeneracy becomes subtle. 

The physics discussed in this work provides another perspective to understand the robustness of edge transport in finite-temperature topological systems \cite{thermal2, thermal3, thermal4}. Indeed, any impurities left on the boundary of a system likely has a nonzero velocity due to the unavoidable thermal fluctuations. As we have learned from this work, different bulk bands and edge-state geometries will lead to different responses to a moving impurity, and hence different responses to temperature effects as well.   

It will be of great interest to investigate how particle-particle interaction or particle quantum statistics would change the qualitative insight developed in this work.  For example, the spectral degeneracy between the initial state on the edge and the bulk states in the co-moving frame would have different consequences for Fermionic bands in electronic systems (where population transition from the edge to the already occupied bulk bands will be suppressed) and Bosonic topological systems, such as photonics systems.



\section{Acknowledgement}
 J.G. acknowledges support by the National Research Foundation, Singapore, through the National Quantum Office, hosted in A*STAR, under its Centre for Quantum Technologies Funding Initiative (S24Q2d0009). J.G. also thanks Gil Refael for suggesting the co-moving frame approach and other helpful discussions during his visit to Singapore.   J.G. also thanks Liujun Zou for stimulating discussions.





\appendix
\section{edge Hamiltonian from continuum model}\label{append_A}
Following Ref.~\cite{append_1,append_2, append_3}, we derive here the edge Hamiltonian used in the main text.  We begin by expanding the QWZ Hamiltonian near the gap-closing point at $k_y=\pi, k_x=\pi$
\begin{equation}
    H(p_x, p_y) \approx - Jp_x \sigma_x - p_y \sigma_y + (m - 2 + \frac{p_x^2+p_y^2}{2})\sigma_z,
\end{equation}
where $p_x = k_x - \pi, p_y = k_y - \pi$. At $m = 2$ and small $p_x, p_y$, this reduces to a massless Dirac Hamiltonian
\begin{equation}
    H(p_x, p_y) \approx J p_x \sigma_x + p_y \sigma_y.
\end{equation}
The overall sign of the above Dirac Hamiltonian has been neglected without loss of generality. We now introduce a spatially varying mass term $m(y) = m_0 \text{sgn}(y)$ as a domain wall and separate the Hamiltonian into
\begin{equation}
    H = H_x + H_y,
\end{equation}
where $H_x = Jp_x \sigma_x$ and $H_y = p_y\sigma_y + m(y)\sigma_z$. Due to translational invariance along $x$, $p_x$ remains a good quantum number. At $p_x = 0$, we solve for zero-energy bound states of
\begin{equation}
    H_y\psi(y) = [-i\partial_y\sigma_y + m(y)\sigma_z]\psi(y).
\end{equation}
Assuming the ansatz $\psi(y) = e^{-\lambda |y|}|\phi_0\rangle$ where $|\phi_0\rangle$ is a spinor, we find that for $y>0$
\begin{equation}\label{append_phi_0}
    [i\lambda \sigma_y + m_0 \sigma_z]| \phi_0\rangle = 0.
\end{equation}
For the solution $|\phi_0\rangle$ to the equation above, and also the eigenstate of the full $H$,  we require that the spinor $|\phi_0\rangle$ be also an eigenstate of $\sigma_x$. This is true if $\lambda = \pm m_0$ and $\ket{\phi_0} = \mqty(1 \\ \dfrac{m_0}{\lambda})$. Now evaluating the full Hamiltonian $H$ on the spinor $\ket{\phi_0}$,  we arrive the following edge state dispersion
\begin{equation}
    H_{\text{edge}} = \bra{\phi_0}H\ket{\phi_0} = \pm J p_x.
\end{equation}
Noting that the edge state stays as an eigenstate of $\sigma_x$, the $y=0$ edge Hamiltonian can then be written as $H_{\rm edge}=Jp_x\sigma_x$.
For cases with $m\ne 2$, the system is now gapped with a massive Dirac Hamiltonian
\begin{equation}
    H(k) = Jp_x\sigma_x +p_y\sigma_y - (m-2)\sigma_z.
\end{equation}
Almost the same reasoning follows, except that the specific relation between $\lambda$ and $m_0$ should be updated. Clearly then, the edge state Hamiltonian at the $y=0$ edge can be described by
\begin{equation}
    H_{\text{edge}} = J p_x \sigma_x.
\end{equation}

\bibliography{apssamp}

\begin{thebibliography}{35}%
\makeatletter
\providecommand \@ifxundefined [1]{%
 \@ifx{#1\undefined}
}%
\providecommand \@ifnum [1]{%
 \ifnum #1\expandafter \@firstoftwo
 \else \expandafter \@secondoftwo
 \fi
}%
\providecommand \@ifx [1]{%
 \ifx #1\expandafter \@firstoftwo
 \else \expandafter \@secondoftwo
 \fi
}%
\providecommand \natexlab [1]{#1}%
\providecommand \enquote  [1]{``#1''}%
\providecommand \bibnamefont  [1]{#1}%
\providecommand \bibfnamefont [1]{#1}%
\providecommand \citenamefont [1]{#1}%
\providecommand \href@noop [0]{\@secondoftwo}%
\providecommand \href [0]{\begingroup \@sanitize@url \@href}%
\providecommand \@href[1]{\@@startlink{#1}\@@href}%
\providecommand \@@href[1]{\endgroup#1\@@endlink}%
\providecommand \@sanitize@url [0]{\catcode `\\12\catcode `\$12\catcode `\&12\catcode `\#12\catcode `\^12\catcode `\_12\catcode `\%12\relax}%
\providecommand \@@startlink[1]{}%
\providecommand \@@endlink[0]{}%
\providecommand \url  [0]{\begingroup\@sanitize@url \@url }%
\providecommand \@url [1]{\endgroup\@href {#1}{\urlprefix }}%
\providecommand \urlprefix  [0]{URL }%
\providecommand \Eprint [0]{\href }%
\providecommand \doibase [0]{https://doi.org/}%
\providecommand \selectlanguage [0]{\@gobble}%
\providecommand \bibinfo  [0]{\@secondoftwo}%
\providecommand \bibfield  [0]{\@secondoftwo}%
\providecommand \translation [1]{[#1]}%
\providecommand \BibitemOpen [0]{}%
\providecommand \bibitemStop [0]{}%
\providecommand \bibitemNoStop [0]{.\EOS\space}%
\providecommand \EOS [0]{\spacefactor3000\relax}%
\providecommand \BibitemShut  [1]{\csname bibitem#1\endcsname}%
\let\auto@bib@innerbib\@empty
\bibitem [{\citenamefont {Thouless}\ \emph {et~al.}(1982)\citenamefont {Thouless}, \citenamefont {Kohmoto}, \citenamefont {Nightingale},\ and\ \citenamefont {den Nijs}}]{TKNN}%
  \BibitemOpen
  \bibfield  {author} {\bibinfo {author} {\bibfnamefont {D.~J.}\ \bibnamefont {Thouless}}, \bibinfo {author} {\bibfnamefont {M.}~\bibnamefont {Kohmoto}}, \bibinfo {author} {\bibfnamefont {M.~P.}\ \bibnamefont {Nightingale}},\ and\ \bibinfo {author} {\bibfnamefont {M.}~\bibnamefont {den Nijs}},\ }\bibfield  {title} {\bibinfo {title} {Quantized hall conductance in a two-dimensional periodic potential},\ }\href {https://doi.org/10.1103/PhysRevLett.49.405} {\bibfield  {journal} {\bibinfo  {journal} {Phys. Rev. Lett.}\ }\textbf {\bibinfo {volume} {49}},\ \bibinfo {pages} {405} (\bibinfo {year} {1982})}\BibitemShut {NoStop}%
\bibitem [{\citenamefont {König}\ \emph {et~al.}(2007)\citenamefont {König}, \citenamefont {Wiedmann}, \citenamefont {Brüne}, \citenamefont {Roth}, \citenamefont {Buhmann}, \citenamefont {Molenkamp}, \citenamefont {Qi},\ and\ \citenamefont {Zhang}}]{zhang_experiment}%
  \BibitemOpen
  \bibfield  {author} {\bibinfo {author} {\bibfnamefont {M.}~\bibnamefont {König}}, \bibinfo {author} {\bibfnamefont {S.}~\bibnamefont {Wiedmann}}, \bibinfo {author} {\bibfnamefont {C.}~\bibnamefont {Brüne}}, \bibinfo {author} {\bibfnamefont {A.}~\bibnamefont {Roth}}, \bibinfo {author} {\bibfnamefont {H.}~\bibnamefont {Buhmann}}, \bibinfo {author} {\bibfnamefont {L.~W.}\ \bibnamefont {Molenkamp}}, \bibinfo {author} {\bibfnamefont {X.-L.}\ \bibnamefont {Qi}},\ and\ \bibinfo {author} {\bibfnamefont {S.-C.}\ \bibnamefont {Zhang}},\ }\bibfield  {title} {\bibinfo {title} {Quantum spin hall insulator state in hgte quantum wells},\ }\href {https://doi.org/10.1126/science.1148047} {\bibfield  {journal} {\bibinfo  {journal} {Science}\ }\textbf {\bibinfo {volume} {318}},\ \bibinfo {pages} {766} (\bibinfo {year} {2007})},\ \Eprint {https://arxiv.org/abs/https://www.science.org/doi/pdf/10.1126/science.1148047} {https://www.science.org/doi/pdf/10.1126/science.1148047} \BibitemShut {NoStop}%
\bibitem [{\citenamefont {Klitzing}\ \emph {et~al.}(1980)\citenamefont {Klitzing}, \citenamefont {Dorda},\ and\ \citenamefont {Pepper}}]{pepper_experiment}%
  \BibitemOpen
  \bibfield  {author} {\bibinfo {author} {\bibfnamefont {K.~v.}\ \bibnamefont {Klitzing}}, \bibinfo {author} {\bibfnamefont {G.}~\bibnamefont {Dorda}},\ and\ \bibinfo {author} {\bibfnamefont {M.}~\bibnamefont {Pepper}},\ }\bibfield  {title} {\bibinfo {title} {New method for high-accuracy determination of the fine-structure constant based on quantized hall resistance},\ }\href {https://doi.org/10.1103/PhysRevLett.45.494} {\bibfield  {journal} {\bibinfo  {journal} {Phys. Rev. Lett.}\ }\textbf {\bibinfo {volume} {45}},\ \bibinfo {pages} {494} (\bibinfo {year} {1980})}\BibitemShut {NoStop}%
\bibitem [{\citenamefont {Asb{\'o}th}\ \emph {et~al.}(2015)\citenamefont {Asb{\'o}th}, \citenamefont {Oroszl{\'a}ny},\ and\ \citenamefont {P{\'a}lyi}}]{asboth}%
  \BibitemOpen
  \bibfield  {author} {\bibinfo {author} {\bibfnamefont {J.~K.}\ \bibnamefont {Asb{\'o}th}}, \bibinfo {author} {\bibfnamefont {L.}~\bibnamefont {Oroszl{\'a}ny}},\ and\ \bibinfo {author} {\bibfnamefont {A.}~\bibnamefont {P{\'a}lyi}},\ }\bibfield  {title} {\bibinfo {title} {A short course on topological insulators: Band-structure topology and edge states in one and two dimensions},\ }\href@noop {} {\bibfield  {journal} {\bibinfo  {journal} {arXiv preprint arXiv:1509.02295}\ } (\bibinfo {year} {2015})}\BibitemShut {NoStop}%
\bibitem [{\citenamefont {Chang}\ \emph {et~al.}(2015)\citenamefont {Chang}, \citenamefont {Zhao}, \citenamefont {Kim}, \citenamefont {Wei}, \citenamefont {Jain}, \citenamefont {Liu}, \citenamefont {Chan},\ and\ \citenamefont {Moodera}}]{chang_spin}%
  \BibitemOpen
  \bibfield  {author} {\bibinfo {author} {\bibfnamefont {C.-Z.}\ \bibnamefont {Chang}}, \bibinfo {author} {\bibfnamefont {W.}~\bibnamefont {Zhao}}, \bibinfo {author} {\bibfnamefont {D.~Y.}\ \bibnamefont {Kim}}, \bibinfo {author} {\bibfnamefont {P.}~\bibnamefont {Wei}}, \bibinfo {author} {\bibfnamefont {J.~K.}\ \bibnamefont {Jain}}, \bibinfo {author} {\bibfnamefont {C.}~\bibnamefont {Liu}}, \bibinfo {author} {\bibfnamefont {M.~H.}\ \bibnamefont {Chan}},\ and\ \bibinfo {author} {\bibfnamefont {J.~S.}\ \bibnamefont {Moodera}},\ }\bibfield  {title} {\bibinfo {title} {Zero-field dissipationless chiral edge transport and the nature of dissipation in the quantum anomalous hall state},\ }\href@noop {} {\bibfield  {journal} {\bibinfo  {journal} {Physical review letters}\ }\textbf {\bibinfo {volume} {115}},\ \bibinfo {pages} {057206} (\bibinfo {year} {2015})}\BibitemShut {NoStop}%
\bibitem [{\citenamefont {Mellnik}\ \emph {et~al.}(2014)\citenamefont {Mellnik}, \citenamefont {Lee}, \citenamefont {Richardella}, \citenamefont {Grab}, \citenamefont {Mintun}, \citenamefont {Fischer}, \citenamefont {Vaezi}, \citenamefont {Manchon}, \citenamefont {Kim}, \citenamefont {Samarth} \emph {et~al.}}]{spin_1}%
  \BibitemOpen
  \bibfield  {author} {\bibinfo {author} {\bibfnamefont {A.~R.}\ \bibnamefont {Mellnik}}, \bibinfo {author} {\bibfnamefont {J.}~\bibnamefont {Lee}}, \bibinfo {author} {\bibfnamefont {A.}~\bibnamefont {Richardella}}, \bibinfo {author} {\bibfnamefont {J.~L.}\ \bibnamefont {Grab}}, \bibinfo {author} {\bibfnamefont {P.~J.}\ \bibnamefont {Mintun}}, \bibinfo {author} {\bibfnamefont {M.~H.}\ \bibnamefont {Fischer}}, \bibinfo {author} {\bibfnamefont {A.}~\bibnamefont {Vaezi}}, \bibinfo {author} {\bibfnamefont {A.}~\bibnamefont {Manchon}}, \bibinfo {author} {\bibfnamefont {E.-A.}\ \bibnamefont {Kim}}, \bibinfo {author} {\bibfnamefont {N.}~\bibnamefont {Samarth}}, \emph {et~al.},\ }\bibfield  {title} {\bibinfo {title} {Spin-transfer torque generated by a topological insulator},\ }\href@noop {} {\bibfield  {journal} {\bibinfo  {journal} {Nature}\ }\textbf {\bibinfo {volume} {511}},\ \bibinfo {pages} {449} (\bibinfo {year} {2014})}\BibitemShut {NoStop}%
\bibitem [{\citenamefont {Shiomi}\ \emph {et~al.}(2014)\citenamefont {Shiomi}, \citenamefont {Nomura}, \citenamefont {Kajiwara}, \citenamefont {Eto}, \citenamefont {Novak}, \citenamefont {Segawa}, \citenamefont {Ando},\ and\ \citenamefont {Saitoh}}]{spin_2}%
  \BibitemOpen
  \bibfield  {author} {\bibinfo {author} {\bibfnamefont {Y.}~\bibnamefont {Shiomi}}, \bibinfo {author} {\bibfnamefont {K.}~\bibnamefont {Nomura}}, \bibinfo {author} {\bibfnamefont {Y.}~\bibnamefont {Kajiwara}}, \bibinfo {author} {\bibfnamefont {K.}~\bibnamefont {Eto}}, \bibinfo {author} {\bibfnamefont {M.}~\bibnamefont {Novak}}, \bibinfo {author} {\bibfnamefont {K.}~\bibnamefont {Segawa}}, \bibinfo {author} {\bibfnamefont {Y.}~\bibnamefont {Ando}},\ and\ \bibinfo {author} {\bibfnamefont {E.}~\bibnamefont {Saitoh}},\ }\bibfield  {title} {\bibinfo {title} {Spin-electricity conversion induced by spin injection into topological insulators},\ }\href@noop {} {\bibfield  {journal} {\bibinfo  {journal} {Physical review letters}\ }\textbf {\bibinfo {volume} {113}},\ \bibinfo {pages} {196601} (\bibinfo {year} {2014})}\BibitemShut {NoStop}%
\bibitem [{\citenamefont {Fan}\ \emph {et~al.}(2014)\citenamefont {Fan}, \citenamefont {Upadhyaya}, \citenamefont {Kou}, \citenamefont {Lang}, \citenamefont {Takei}, \citenamefont {Wang}, \citenamefont {Tang}, \citenamefont {He}, \citenamefont {Chang}, \citenamefont {Montazeri} \emph {et~al.}}]{spin_3}%
  \BibitemOpen
  \bibfield  {author} {\bibinfo {author} {\bibfnamefont {Y.}~\bibnamefont {Fan}}, \bibinfo {author} {\bibfnamefont {P.}~\bibnamefont {Upadhyaya}}, \bibinfo {author} {\bibfnamefont {X.}~\bibnamefont {Kou}}, \bibinfo {author} {\bibfnamefont {M.}~\bibnamefont {Lang}}, \bibinfo {author} {\bibfnamefont {S.}~\bibnamefont {Takei}}, \bibinfo {author} {\bibfnamefont {Z.}~\bibnamefont {Wang}}, \bibinfo {author} {\bibfnamefont {J.}~\bibnamefont {Tang}}, \bibinfo {author} {\bibfnamefont {L.}~\bibnamefont {He}}, \bibinfo {author} {\bibfnamefont {L.-T.}\ \bibnamefont {Chang}}, \bibinfo {author} {\bibfnamefont {M.}~\bibnamefont {Montazeri}}, \emph {et~al.},\ }\bibfield  {title} {\bibinfo {title} {Magnetization switching through giant spin--orbit torque in a magnetically doped topological insulator heterostructure},\ }\href@noop {} {\bibfield  {journal} {\bibinfo  {journal} {Nature materials}\ }\textbf {\bibinfo {volume} {13}},\ \bibinfo {pages} {699} (\bibinfo {year} {2014})}\BibitemShut {NoStop}%
\bibitem [{\citenamefont {Guo}\ \emph {et~al.}(2021)\citenamefont {Guo}, \citenamefont {Zhang}, \citenamefont {Song}, \citenamefont {Jiang},\ and\ \citenamefont {Chen}}]{sensing_1}%
  \BibitemOpen
  \bibfield  {author} {\bibinfo {author} {\bibfnamefont {Z.}~\bibnamefont {Guo}}, \bibinfo {author} {\bibfnamefont {T.}~\bibnamefont {Zhang}}, \bibinfo {author} {\bibfnamefont {J.}~\bibnamefont {Song}}, \bibinfo {author} {\bibfnamefont {H.}~\bibnamefont {Jiang}},\ and\ \bibinfo {author} {\bibfnamefont {H.}~\bibnamefont {Chen}},\ }\bibfield  {title} {\bibinfo {title} {Sensitivity of topological edge states in a non-hermitian dimer chain},\ }\href@noop {} {\bibfield  {journal} {\bibinfo  {journal} {Photonics Research}\ }\textbf {\bibinfo {volume} {9}},\ \bibinfo {pages} {574} (\bibinfo {year} {2021})}\BibitemShut {NoStop}%
\bibitem [{\citenamefont {Sarkar}\ \emph {et~al.}(2022)\citenamefont {Sarkar}, \citenamefont {Mukhopadhyay}, \citenamefont {Alase},\ and\ \citenamefont {Bayat}}]{sensing2}%
  \BibitemOpen
  \bibfield  {author} {\bibinfo {author} {\bibfnamefont {S.}~\bibnamefont {Sarkar}}, \bibinfo {author} {\bibfnamefont {C.}~\bibnamefont {Mukhopadhyay}}, \bibinfo {author} {\bibfnamefont {A.}~\bibnamefont {Alase}},\ and\ \bibinfo {author} {\bibfnamefont {A.}~\bibnamefont {Bayat}},\ }\bibfield  {title} {\bibinfo {title} {Free-fermionic topological quantum sensors},\ }\href {https://doi.org/10.1103/PhysRevLett.129.090503} {\bibfield  {journal} {\bibinfo  {journal} {Phys. Rev. Lett.}\ }\textbf {\bibinfo {volume} {129}},\ \bibinfo {pages} {090503} (\bibinfo {year} {2022})}\BibitemShut {NoStop}%
\bibitem [{\citenamefont {Li}\ \emph {et~al.}(2009)\citenamefont {Li}, \citenamefont {Chu}, \citenamefont {Jain},\ and\ \citenamefont {Shen}}]{TAI_1}%
  \BibitemOpen
  \bibfield  {author} {\bibinfo {author} {\bibfnamefont {J.}~\bibnamefont {Li}}, \bibinfo {author} {\bibfnamefont {R.-L.}\ \bibnamefont {Chu}}, \bibinfo {author} {\bibfnamefont {J.~K.}\ \bibnamefont {Jain}},\ and\ \bibinfo {author} {\bibfnamefont {S.-Q.}\ \bibnamefont {Shen}},\ }\bibfield  {title} {\bibinfo {title} {Topological anderson insulator},\ }\href@noop {} {\bibfield  {journal} {\bibinfo  {journal} {Physical review letters}\ }\textbf {\bibinfo {volume} {102}},\ \bibinfo {pages} {136806} (\bibinfo {year} {2009})}\BibitemShut {NoStop}%
\bibitem [{\citenamefont {Groth}\ \emph {et~al.}(2009)\citenamefont {Groth}, \citenamefont {Wimmer}, \citenamefont {Akhmerov}, \citenamefont {Tworzyd{\l}o},\ and\ \citenamefont {Beenakker}}]{TAI_2}%
  \BibitemOpen
  \bibfield  {author} {\bibinfo {author} {\bibfnamefont {C.}~\bibnamefont {Groth}}, \bibinfo {author} {\bibfnamefont {M.}~\bibnamefont {Wimmer}}, \bibinfo {author} {\bibfnamefont {A.}~\bibnamefont {Akhmerov}}, \bibinfo {author} {\bibfnamefont {J.}~\bibnamefont {Tworzyd{\l}o}},\ and\ \bibinfo {author} {\bibfnamefont {C.}~\bibnamefont {Beenakker}},\ }\bibfield  {title} {\bibinfo {title} {Theory of the topological anderson insulator},\ }\href@noop {} {\bibfield  {journal} {\bibinfo  {journal} {Physical review letters}\ }\textbf {\bibinfo {volume} {103}},\ \bibinfo {pages} {196805} (\bibinfo {year} {2009})}\BibitemShut {NoStop}%
\bibitem [{\citenamefont {St{\"u}tzer}\ \emph {et~al.}(2018)\citenamefont {St{\"u}tzer}, \citenamefont {Plotnik}, \citenamefont {Lumer}, \citenamefont {Titum}, \citenamefont {Lindner}, \citenamefont {Segev}, \citenamefont {Rechtsman},\ and\ \citenamefont {Szameit}}]{TAI_3}%
  \BibitemOpen
  \bibfield  {author} {\bibinfo {author} {\bibfnamefont {S.}~\bibnamefont {St{\"u}tzer}}, \bibinfo {author} {\bibfnamefont {Y.}~\bibnamefont {Plotnik}}, \bibinfo {author} {\bibfnamefont {Y.}~\bibnamefont {Lumer}}, \bibinfo {author} {\bibfnamefont {P.}~\bibnamefont {Titum}}, \bibinfo {author} {\bibfnamefont {N.~H.}\ \bibnamefont {Lindner}}, \bibinfo {author} {\bibfnamefont {M.}~\bibnamefont {Segev}}, \bibinfo {author} {\bibfnamefont {M.~C.}\ \bibnamefont {Rechtsman}},\ and\ \bibinfo {author} {\bibfnamefont {A.}~\bibnamefont {Szameit}},\ }\bibfield  {title} {\bibinfo {title} {Photonic topological anderson insulators},\ }\href@noop {} {\bibfield  {journal} {\bibinfo  {journal} {Nature}\ }\textbf {\bibinfo {volume} {560}},\ \bibinfo {pages} {461} (\bibinfo {year} {2018})}\BibitemShut {NoStop}%
\bibitem [{\citenamefont {Guo}\ \emph {et~al.}(2010)\citenamefont {Guo}, \citenamefont {Rosenberg}, \citenamefont {Refael},\ and\ \citenamefont {Franz}}]{TAI_4}%
  \BibitemOpen
  \bibfield  {author} {\bibinfo {author} {\bibfnamefont {H.-M.}\ \bibnamefont {Guo}}, \bibinfo {author} {\bibfnamefont {G.}~\bibnamefont {Rosenberg}}, \bibinfo {author} {\bibfnamefont {G.}~\bibnamefont {Refael}},\ and\ \bibinfo {author} {\bibfnamefont {M.}~\bibnamefont {Franz}},\ }\bibfield  {title} {\bibinfo {title} {Topological anderson insulator in three dimensions},\ }\href@noop {} {\bibfield  {journal} {\bibinfo  {journal} {Physical review letters}\ }\textbf {\bibinfo {volume} {105}},\ \bibinfo {pages} {216601} (\bibinfo {year} {2010})}\BibitemShut {NoStop}%
\bibitem [{\citenamefont {Gneiting}\ and\ \citenamefont {Nori}(2017)}]{dephasing}%
  \BibitemOpen
  \bibfield  {author} {\bibinfo {author} {\bibfnamefont {C.}~\bibnamefont {Gneiting}}\ and\ \bibinfo {author} {\bibfnamefont {F.}~\bibnamefont {Nori}},\ }\bibfield  {title} {\bibinfo {title} {Disorder-induced dephasing in backscattering-free quantum transport},\ }\href {https://doi.org/10.1103/PhysRevLett.119.176802} {\bibfield  {journal} {\bibinfo  {journal} {Phys. Rev. Lett.}\ }\textbf {\bibinfo {volume} {119}},\ \bibinfo {pages} {176802} (\bibinfo {year} {2017})}\BibitemShut {NoStop}%
\bibitem [{\citenamefont {Gneiting}\ \emph {et~al.}(2019)\citenamefont {Gneiting}, \citenamefont {Leykam},\ and\ \citenamefont {Nori}}]{entanglement_transport}%
  \BibitemOpen
  \bibfield  {author} {\bibinfo {author} {\bibfnamefont {C.}~\bibnamefont {Gneiting}}, \bibinfo {author} {\bibfnamefont {D.}~\bibnamefont {Leykam}},\ and\ \bibinfo {author} {\bibfnamefont {F.}~\bibnamefont {Nori}},\ }\bibfield  {title} {\bibinfo {title} {Disorder-robust entanglement transport},\ }\href@noop {} {\bibfield  {journal} {\bibinfo  {journal} {Physical review letters}\ }\textbf {\bibinfo {volume} {122}},\ \bibinfo {pages} {066601} (\bibinfo {year} {2019})}\BibitemShut {NoStop}%
\bibitem [{\citenamefont {Hsu}\ \emph {et~al.}(2021)\citenamefont {Hsu}, \citenamefont {Stano}, \citenamefont {Klinovaja},\ and\ \citenamefont {Loss}}]{helical_liquid}%
  \BibitemOpen
  \bibfield  {author} {\bibinfo {author} {\bibfnamefont {C.-H.}\ \bibnamefont {Hsu}}, \bibinfo {author} {\bibfnamefont {P.}~\bibnamefont {Stano}}, \bibinfo {author} {\bibfnamefont {J.}~\bibnamefont {Klinovaja}},\ and\ \bibinfo {author} {\bibfnamefont {D.}~\bibnamefont {Loss}},\ }\bibfield  {title} {\bibinfo {title} {Helical liquids in semiconductors},\ }\href@noop {} {\bibfield  {journal} {\bibinfo  {journal} {Semiconductor Science and Technology}\ }\textbf {\bibinfo {volume} {36}},\ \bibinfo {pages} {123003} (\bibinfo {year} {2021})}\BibitemShut {NoStop}%
\bibitem [{\citenamefont {V\"ayrynen}\ \emph {et~al.}(2018{\natexlab{a}})\citenamefont {V\"ayrynen}, \citenamefont {Pikulin},\ and\ \citenamefont {Alicea}}]{noise1}%
  \BibitemOpen
  \bibfield  {author} {\bibinfo {author} {\bibfnamefont {J.~I.}\ \bibnamefont {V\"ayrynen}}, \bibinfo {author} {\bibfnamefont {D.~I.}\ \bibnamefont {Pikulin}},\ and\ \bibinfo {author} {\bibfnamefont {J.}~\bibnamefont {Alicea}},\ }\bibfield  {title} {\bibinfo {title} {Noise-induced backscattering in a quantum spin hall edge},\ }\href {https://doi.org/10.1103/PhysRevLett.121.106601} {\bibfield  {journal} {\bibinfo  {journal} {Phys. Rev. Lett.}\ }\textbf {\bibinfo {volume} {121}},\ \bibinfo {pages} {106601} (\bibinfo {year} {2018}{\natexlab{a}})}\BibitemShut {NoStop}%
\bibitem [{\citenamefont {Altshuler}\ \emph {et~al.}(2013)\citenamefont {Altshuler}, \citenamefont {Aleiner},\ and\ \citenamefont {Yudson}}]{noise2}%
  \BibitemOpen
  \bibfield  {author} {\bibinfo {author} {\bibfnamefont {B.~L.}\ \bibnamefont {Altshuler}}, \bibinfo {author} {\bibfnamefont {I.~L.}\ \bibnamefont {Aleiner}},\ and\ \bibinfo {author} {\bibfnamefont {V.~I.}\ \bibnamefont {Yudson}},\ }\bibfield  {title} {\bibinfo {title} {Localization at the edge of a 2d topological insulator by kondo impurities with random anisotropies},\ }\href {https://doi.org/10.1103/PhysRevLett.111.086401} {\bibfield  {journal} {\bibinfo  {journal} {Phys. Rev. Lett.}\ }\textbf {\bibinfo {volume} {111}},\ \bibinfo {pages} {086401} (\bibinfo {year} {2013})}\BibitemShut {NoStop}%
\bibitem [{\citenamefont {Valiente}(2019)}]{moving_1}%
  \BibitemOpen
  \bibfield  {author} {\bibinfo {author} {\bibfnamefont {M.}~\bibnamefont {Valiente}},\ }\bibfield  {title} {\bibinfo {title} {Flat band of topological states bound to a mobile impurity},\ }\href@noop {} {\bibfield  {journal} {\bibinfo  {journal} {arXiv preprint arXiv:1907.08215}\ } (\bibinfo {year} {2019})}\BibitemShut {NoStop}%
\bibitem [{\citenamefont {Karzig}\ \emph {et~al.}(2013)\citenamefont {Karzig}, \citenamefont {Refael},\ and\ \citenamefont {von Oppen}}]{moving_2}%
  \BibitemOpen
  \bibfield  {author} {\bibinfo {author} {\bibfnamefont {T.}~\bibnamefont {Karzig}}, \bibinfo {author} {\bibfnamefont {G.}~\bibnamefont {Refael}},\ and\ \bibinfo {author} {\bibfnamefont {F.}~\bibnamefont {von Oppen}},\ }\bibfield  {title} {\bibinfo {title} {Boosting majorana zero modes},\ }\href@noop {} {\bibfield  {journal} {\bibinfo  {journal} {Physical Review X}\ }\textbf {\bibinfo {volume} {3}},\ \bibinfo {pages} {041017} (\bibinfo {year} {2013})}\BibitemShut {NoStop}%
\bibitem [{\citenamefont {Wozny}\ \emph {et~al.}(2021)\citenamefont {Wozny}, \citenamefont {Leijnse},\ and\ \citenamefont {Erlingsson}}]{moving_3}%
  \BibitemOpen
  \bibfield  {author} {\bibinfo {author} {\bibfnamefont {S.}~\bibnamefont {Wozny}}, \bibinfo {author} {\bibfnamefont {M.}~\bibnamefont {Leijnse}},\ and\ \bibinfo {author} {\bibfnamefont {S.~I.}\ \bibnamefont {Erlingsson}},\ }\bibfield  {title} {\bibinfo {title} {Dynamic impurities in two-dimensional topological-insulator edge states},\ }\href@noop {} {\bibfield  {journal} {\bibinfo  {journal} {Physical Review B}\ }\textbf {\bibinfo {volume} {104}},\ \bibinfo {pages} {205418} (\bibinfo {year} {2021})}\BibitemShut {NoStop}%
\bibitem [{\citenamefont {Feit}\ \emph {et~al.}(1982{\natexlab{a}})\citenamefont {Feit}, \citenamefont {Fleck~Jr},\ and\ \citenamefont {Steiger}}]{trotter_1}%
  \BibitemOpen
  \bibfield  {author} {\bibinfo {author} {\bibfnamefont {M.}~\bibnamefont {Feit}}, \bibinfo {author} {\bibfnamefont {J.}~\bibnamefont {Fleck~Jr}},\ and\ \bibinfo {author} {\bibfnamefont {A.}~\bibnamefont {Steiger}},\ }\bibfield  {title} {\bibinfo {title} {Solution of the schr{\"o}dinger equation by a spectral method},\ }\href@noop {} {\bibfield  {journal} {\bibinfo  {journal} {Journal of Computational Physics}\ }\textbf {\bibinfo {volume} {47}},\ \bibinfo {pages} {412} (\bibinfo {year} {1982}{\natexlab{a}})}\BibitemShut {NoStop}%
\bibitem [{\citenamefont {Feit}\ \emph {et~al.}(1982{\natexlab{b}})\citenamefont {Feit}, \citenamefont {Fleck~Jr},\ and\ \citenamefont {Steiger}}]{trotter_2}%
  \BibitemOpen
  \bibfield  {author} {\bibinfo {author} {\bibfnamefont {M.}~\bibnamefont {Feit}}, \bibinfo {author} {\bibfnamefont {J.}~\bibnamefont {Fleck~Jr}},\ and\ \bibinfo {author} {\bibfnamefont {A.}~\bibnamefont {Steiger}},\ }\bibfield  {title} {\bibinfo {title} {Solution of the schr{\"o}dinger equation by a spectral method},\ }\href@noop {} {\bibfield  {journal} {\bibinfo  {journal} {Journal of Computational Physics}\ }\textbf {\bibinfo {volume} {47}},\ \bibinfo {pages} {412} (\bibinfo {year} {1982}{\natexlab{b}})}\BibitemShut {NoStop}%
\bibitem [{\citenamefont {V{\"a}yrynen}\ \emph {et~al.}(2013)\citenamefont {V{\"a}yrynen}, \citenamefont {Goldstein},\ and\ \citenamefont {Glazman}}]{group_1}%
  \BibitemOpen
  \bibfield  {author} {\bibinfo {author} {\bibfnamefont {J.~I.}\ \bibnamefont {V{\"a}yrynen}}, \bibinfo {author} {\bibfnamefont {M.}~\bibnamefont {Goldstein}},\ and\ \bibinfo {author} {\bibfnamefont {L.~I.}\ \bibnamefont {Glazman}},\ }\bibfield  {title} {\bibinfo {title} {Helical edge resistance introduced by charge puddles},\ }\href@noop {} {\bibfield  {journal} {\bibinfo  {journal} {Physical Review Letters}\ }\textbf {\bibinfo {volume} {110}},\ \bibinfo {pages} {216402} (\bibinfo {year} {2013})}\BibitemShut {NoStop}%
\bibitem [{\citenamefont {Maciejko}\ \emph {et~al.}(2009)\citenamefont {Maciejko}, \citenamefont {Liu}, \citenamefont {Oreg}, \citenamefont {Qi}, \citenamefont {Wu},\ and\ \citenamefont {Zhang}}]{group_2}%
  \BibitemOpen
  \bibfield  {author} {\bibinfo {author} {\bibfnamefont {J.}~\bibnamefont {Maciejko}}, \bibinfo {author} {\bibfnamefont {C.}~\bibnamefont {Liu}}, \bibinfo {author} {\bibfnamefont {Y.}~\bibnamefont {Oreg}}, \bibinfo {author} {\bibfnamefont {X.-L.}\ \bibnamefont {Qi}}, \bibinfo {author} {\bibfnamefont {C.}~\bibnamefont {Wu}},\ and\ \bibinfo {author} {\bibfnamefont {S.-C.}\ \bibnamefont {Zhang}},\ }\bibfield  {title} {\bibinfo {title} {Kondo effect in the helical edge liquid of the quantum spin hall state},\ }\href@noop {} {\bibfield  {journal} {\bibinfo  {journal} {Physical review letters}\ }\textbf {\bibinfo {volume} {102}},\ \bibinfo {pages} {256803} (\bibinfo {year} {2009})}\BibitemShut {NoStop}%
\bibitem [{\citenamefont {Bernevig}\ \emph {et~al.}(2006)\citenamefont {Bernevig}, \citenamefont {Hughes},\ and\ \citenamefont {Zhang}}]{BHZ}%
  \BibitemOpen
  \bibfield  {author} {\bibinfo {author} {\bibfnamefont {B.~A.}\ \bibnamefont {Bernevig}}, \bibinfo {author} {\bibfnamefont {T.~L.}\ \bibnamefont {Hughes}},\ and\ \bibinfo {author} {\bibfnamefont {S.-C.}\ \bibnamefont {Zhang}},\ }\bibfield  {title} {\bibinfo {title} {Quantum spin hall effect and topological phase transition in hgte quantum wells},\ }\href@noop {} {\bibfield  {journal} {\bibinfo  {journal} {science}\ }\textbf {\bibinfo {volume} {314}},\ \bibinfo {pages} {1757} (\bibinfo {year} {2006})}\BibitemShut {NoStop}%
\bibitem [{\citenamefont {Umer}\ \emph {et~al.}(2020)\citenamefont {Umer}, \citenamefont {Bomantara},\ and\ \citenamefont {Gong}}]{Umer}%
  \BibitemOpen
  \bibfield  {author} {\bibinfo {author} {\bibfnamefont {M.}~\bibnamefont {Umer}}, \bibinfo {author} {\bibfnamefont {R.~W.}\ \bibnamefont {Bomantara}},\ and\ \bibinfo {author} {\bibfnamefont {J.}~\bibnamefont {Gong}},\ }\bibfield  {title} {\bibinfo {title} {Counterpropagating edge states in floquet topological insulating phases},\ }\href {https://doi.org/10.1103/PhysRevB.101.235438} {\bibfield  {journal} {\bibinfo  {journal} {Phys. Rev. B}\ }\textbf {\bibinfo {volume} {101}},\ \bibinfo {pages} {235438} (\bibinfo {year} {2020})}\BibitemShut {NoStop}%
\bibitem [{\citenamefont {Zhou}\ and\ \citenamefont {Gong}(2018)}]{recipe}%
  \BibitemOpen
  \bibfield  {author} {\bibinfo {author} {\bibfnamefont {L.}~\bibnamefont {Zhou}}\ and\ \bibinfo {author} {\bibfnamefont {J.}~\bibnamefont {Gong}},\ }\bibfield  {title} {\bibinfo {title} {Recipe for creating an arbitrary number of floquet chiral edge states},\ }\href@noop {} {\bibfield  {journal} {\bibinfo  {journal} {Physical Review B}\ }\textbf {\bibinfo {volume} {97}},\ \bibinfo {pages} {245430} (\bibinfo {year} {2018})}\BibitemShut {NoStop}%
\bibitem [{\citenamefont {Roberts}\ \emph {et~al.}(2017)\citenamefont {Roberts}, \citenamefont {Yoshida}, \citenamefont {Kubica},\ and\ \citenamefont {Bartlett}}]{thermal2}%
  \BibitemOpen
  \bibfield  {author} {\bibinfo {author} {\bibfnamefont {S.}~\bibnamefont {Roberts}}, \bibinfo {author} {\bibfnamefont {B.}~\bibnamefont {Yoshida}}, \bibinfo {author} {\bibfnamefont {A.}~\bibnamefont {Kubica}},\ and\ \bibinfo {author} {\bibfnamefont {S.~D.}\ \bibnamefont {Bartlett}},\ }\bibfield  {title} {\bibinfo {title} {Symmetry-protected topological order at nonzero temperature},\ }\href {https://doi.org/10.1103/PhysRevA.96.022306} {\bibfield  {journal} {\bibinfo  {journal} {Phys. Rev. A}\ }\textbf {\bibinfo {volume} {96}},\ \bibinfo {pages} {022306} (\bibinfo {year} {2017})}\BibitemShut {NoStop}%
\bibitem [{\citenamefont {Hastings}(2011)}]{thermal3}%
  \BibitemOpen
  \bibfield  {author} {\bibinfo {author} {\bibfnamefont {M.~B.}\ \bibnamefont {Hastings}},\ }\bibfield  {title} {\bibinfo {title} {Topological order at nonzero temperature},\ }\href {https://doi.org/10.1103/PhysRevLett.107.210501} {\bibfield  {journal} {\bibinfo  {journal} {Phys. Rev. Lett.}\ }\textbf {\bibinfo {volume} {107}},\ \bibinfo {pages} {210501} (\bibinfo {year} {2011})}\BibitemShut {NoStop}%
\bibitem [{\citenamefont {V\"ayrynen}\ \emph {et~al.}(2018{\natexlab{b}})\citenamefont {V\"ayrynen}, \citenamefont {Pikulin},\ and\ \citenamefont {Alicea}}]{thermal4}%
  \BibitemOpen
  \bibfield  {author} {\bibinfo {author} {\bibfnamefont {J.~I.}\ \bibnamefont {V\"ayrynen}}, \bibinfo {author} {\bibfnamefont {D.~I.}\ \bibnamefont {Pikulin}},\ and\ \bibinfo {author} {\bibfnamefont {J.}~\bibnamefont {Alicea}},\ }\bibfield  {title} {\bibinfo {title} {Noise-induced backscattering in a quantum spin hall edge},\ }\href {https://doi.org/10.1103/PhysRevLett.121.106601} {\bibfield  {journal} {\bibinfo  {journal} {Phys. Rev. Lett.}\ }\textbf {\bibinfo {volume} {121}},\ \bibinfo {pages} {106601} (\bibinfo {year} {2018}{\natexlab{b}})}\BibitemShut {NoStop}%
\bibitem [{\citenamefont {Jackiw}\ and\ \citenamefont {Rebbi}(1976)}]{append_1}%
  \BibitemOpen
  \bibfield  {author} {\bibinfo {author} {\bibfnamefont {R.}~\bibnamefont {Jackiw}}\ and\ \bibinfo {author} {\bibfnamefont {C.}~\bibnamefont {Rebbi}},\ }\bibfield  {title} {\bibinfo {title} {Solitons with fermion number $1/2$},\ }\href@noop {} {\bibfield  {journal} {\bibinfo  {journal} {Physical Review D}\ }\textbf {\bibinfo {volume} {13}},\ \bibinfo {pages} {3398} (\bibinfo {year} {1976})}\BibitemShut {NoStop}%
\bibitem [{\citenamefont {Qi}\ and\ \citenamefont {Zhang}(2011)}]{append_2}%
  \BibitemOpen
  \bibfield  {author} {\bibinfo {author} {\bibfnamefont {X.-L.}\ \bibnamefont {Qi}}\ and\ \bibinfo {author} {\bibfnamefont {S.-C.}\ \bibnamefont {Zhang}},\ }\bibfield  {title} {\bibinfo {title} {Topological insulators and superconductors},\ }\href@noop {} {\bibfield  {journal} {\bibinfo  {journal} {Reviews of modern physics}\ }\textbf {\bibinfo {volume} {83}},\ \bibinfo {pages} {1057} (\bibinfo {year} {2011})}\BibitemShut {NoStop}%
\bibitem [{\citenamefont {K{\"o}nig}\ \emph {et~al.}(2008)\citenamefont {K{\"o}nig}, \citenamefont {Buhmann}, \citenamefont {W.~Molenkamp}, \citenamefont {Hughes}, \citenamefont {Liu}, \citenamefont {Qi},\ and\ \citenamefont {Zhang}}]{append_3}%
  \BibitemOpen
  \bibfield  {author} {\bibinfo {author} {\bibfnamefont {M.}~\bibnamefont {K{\"o}nig}}, \bibinfo {author} {\bibfnamefont {H.}~\bibnamefont {Buhmann}}, \bibinfo {author} {\bibfnamefont {L.}~\bibnamefont {W.~Molenkamp}}, \bibinfo {author} {\bibfnamefont {T.}~\bibnamefont {Hughes}}, \bibinfo {author} {\bibfnamefont {C.-X.}\ \bibnamefont {Liu}}, \bibinfo {author} {\bibfnamefont {X.-L.}\ \bibnamefont {Qi}},\ and\ \bibinfo {author} {\bibfnamefont {S.-C.}\ \bibnamefont {Zhang}},\ }\bibfield  {title} {\bibinfo {title} {The quantum spin hall effect: theory and experiment},\ }\href@noop {} {\bibfield  {journal} {\bibinfo  {journal} {Journal of the Physical Society of Japan}\ }\textbf {\bibinfo {volume} {77}},\ \bibinfo {pages} {031007} (\bibinfo {year} {2008})}\BibitemShut {NoStop}%
\end{thebibliography}%

\end{document}